\documentstyle[aps,prb,preprint]{revtex}
\newcommand{\e}{{\mathrm e}}
\renewcommand{\d}{{\mathrm d}}
\newcommand{\bk}{{\mathbf k}}
\newcommand{\bth}{{\mathbf\theta}}
\newcommand{\ve}{\varepsilon}
\newcommand{\D}{{\cal D}}

\begin{document}
\draft
\preprint{preprint 95/4}
\title{Renormalization group approach to anisotropic superconductors
at finite temperature.}
\author{Fabio Siringo, Giuseppe G.N. Angilella and Renato Pucci}
\address{Dipartimento di Fisica dell'Universit\`a di Catania,\\
Corso Italia 57, I 95129 Catania, Italy}
\date{\today}
\maketitle
\begin{abstract}
A renormalization group (RG) analysis of the superconductive instability
of an anisotropic fermionic system is developed at a finite temperature.
The method appears a natural generalization of
Shankar's approach to
interacting fermions and of Weinberg's discussion about
anisotropic superconductors at $T=0$.
The need of such an extension is fully justified by the effectiveness of the
RG at the critical point. Moreover the relationship between the RG and
a mean-field approach is clarified, and
a scale-invariant gap equation is discussed
at a renormalization level in terms
of the eigenfunctions of the interaction potential, regarded as the kernel
of an integral operator on the Fermi surface. At the critical point, the
gap function is expressed by a single eigenfunction and no symmetry mixing is
allowed. As an illustration of the method we discuss an anisotropic
tight-binding model for some classes of high $T_c$ cuprate superconductors,
exhibiting a layered structure.
Some indications on the nature of the pairing interaction emerge
from a comparison of the model predictions with the experimental
data.
\end{abstract}
\pacs{PACS numbers: 11.10.Hi, 64.60.Ak, 74.20.-z, 74.72.-h}
\date{\today}

\section{Introduction}

In a recent review,~\cite{Shankar} Shankar unified a large variety
of physical phenomena and their conventional descriptions, such as
Landau theory of Fermi liquids, charge density waves and nesting in the
Fermi surface, superconductive instabilities and mean-field theories
\emph{etc.,} under the scheme of a renormalization group
(RG) approach, with the aim of a better understanding of the general
nature of the physics involved in such processes. Central to
that approach is the idea that the instability towards a phase transition
in an interacting fermionic system may be discussed  in a very general
fashion, using the RG technique, which involves
the cut-off and the rescaling of
the energies and the momenta close to the Fermi surface. For instance,
a charge density wave or a superconducting condensate are recovered owing
to very general symmetry properties of the single particle dispersion
relation, \emph{i.e.,} the existence of a nesting vector and the validity
of time reversal symmetry, respectively.

In an almost contemporary work,~\cite{Weinberg} Weinberg derived a RG flow
equation for superconductors whose Fermi surface satisfies just time reversal
invariance, employing the standard field-theoretic scheme. In particular,
Weinberg's work doesn't claim for a spherical Fermi surface,
and therefore shows itself more suitable for taking into account anisotropic
materials such as the high $T_c$ layered cuprates.

An extension to a non-zero temperature of the RG analysis is called for,
at least for two major reasons: \emph{i)}
the RG approach is more effective around
the critical point, which is generally located at a finite temperature $T_c$;
\emph{ii)}
in order to make contact with the conventional mean-field approach and with
the experimental data, the RG predictions would be required at a finite
temperature too. Such an extension is straightforward, from a technical
point of view, and even for a generic anisotropic system, the flow equations
are decoupled in terms of the eigenvalues of an integral operator $\hat{V}$,
whose kernel is the marginal pairing coupling evaluated at the Fermi
surface.~\cite{Weinberg}
The critical temperature $T_c$ is then found out to be a
simple function of the most negative eigenvalue, which determines the
transition point. The standard {\sc bcs} expression is recovered for a
spherical Fermi surface with a constant pairing coupling, whereas
for a generic rotationally invariant system the eigenfunctions reduce to
spherical waves, since then the angular momentum and $\hat{V}$ are mutually
commuting operators.

In order to make contact with the standard mean-field approximation, we recover
again the same analytical expression for $T_c$, though following
a quite different path, starting from the usual gap equation. The occurrence in
such an expression of just one eigenvalue is a token of the existence
of a leading part in the pairing coupling, which leads the transition. In fact,
the expansion of the integral operator $\hat{V}$ in terms of its eigenfunctions
allows one to determine its relevant part as its projection
over the eigenfunction belonging to the most negative eigenvalue. In other
words, close to the critical point, the interaction operator $\hat{V}$ acts on
the one-dimensional subspace generated by just one eigenfunction.
All the physical quantities depend on this eigenfunction,
which entirely determines even the symmetry pattern of the
gap function. As a consequence,
the occurrence of any symmetry mixing in the gap function (such as $s$-$d$
wave mixing) is ruled out at the critical point, since in that limit the gap
function reduces to a single eigenfunction with a fixed symmetry (provided
that the most negative eigenvalue is not degenerate).

Even far from the critical point, the expansion of $\hat{V}$ in terms of
its eigenfunctions may be inserted in the gap equation, thus providing
a scale invariant relation, which is of some utility especially when dealing
with separable potentials, when the number of eigenvalues is finite and so
is the expansion in terms of the eigenfunctions.
The resulting set of coupled non-linear equations always admits symmetric
solutions, \emph{i.e.} solutions for the gap function which share the eventual
symmetry of the physical
system. However, far from the critical point the equations
are highly non-linear, so that the uniqueness of the symmetric solutions is not
guaranteed, and broken-symmetry solutions are not forbidden.

Such general aspects of the solutions, as their symmetries
near the critical point, are more easily discussed
by an RG approach than within a mean-field approximation.
This feature, together with a major handiness in dealing
with the numerical cases, provides the RG approach with
more appeal than the mean-field approximation, without
spoiling their equivalence.

A nice illustration of the method is provided by a simple tight-binding
model, recently proposed by Spathis \emph{et al.}~\cite{Schneider,Spathis}
for the single-particle
dispersion relation of a layered high $T_c$ superconductor, namely
Bi$_2$Sr$_2$CaCu$_2$O$_{8+\delta}$ (BSCCO). In order to compare our results
with those already obtained with a conventional {\sc bcs}
procedure,~\cite{Schneider,Spathis} the same
dispersion relation is exactly employed, where rotationally non-invariance
accounts for the structural anisotropy in BSCCO, and two
singlet pairing couplings
are considered, namely an on-site and a nearest-neighbour
interaction.

In particular, the
characteristic dependence of $T_c$ \emph{versus} the carrier concentration $N$
is recovered,~\cite{Spathis} although it seems to be a mere effect of the
peaked quasi-bidimensional density of states. Nonetheless, we observe the
best qualitative agreement with the experimental data when the two couplings
have opposite signs, and precisely in the case of an attractive on-site
interaction (negative Hubbard~\cite{Micnas} $U$) and an inter-site
nearest-neighbour repulsion.

Moreover, while a nearest-neighbour attraction always yields a
superconducting ground state at $T=0$, for any value of the chemical
potential $\mu$ (though the critical temperature may be extremely low), an
on-site attraction is cancelled out by the presence of a nearest-neighbour
repulsion, for a quite large range of values for the chemical potential.

{}From a physical point of view, a negative Hubbard $U$ could be the effect
of a very short ranged coupling interaction, while a nearest-neighbour
repulsion could be justified by the Coulomb long-range potential.

We observe, however, that such comments on the nature of the pairing
interaction are far from being conclusive, and a larger number of pairing
couplings should be retained in the model potential for a full comparison
with the experimental data. From this point of view, the method is easily
implemented, since adding any other coupling merely implies an increase
in the space dimension of the integral operator $\hat{V}$.

The paper is organized as follows. In Section~II we outline the RG approach
at a finite temperature for a generic anisotropic fermionic system.
In Section~III very
similar results are recovered by use of a standard gap equation, and the
possible predictions of the RG approach about the structure and symmetry of
the gap function are discussed. Later in Section~IV, as an illustration of
the method, we consider a tight-binding model for the single-particle
dispersion relation in layered high $T_c$ superconductors, and finally
in Section~V the numerical results are discussed.

\section{RG flow at a finite temperature}

Shankar's~\cite{Shankar} RG approach to the superconductive instability
properties of an interacting fermionic system is here generalized to a
finite temperature and for a generic anisotropic Fermi surface. No other
special symmetry is assumed than time-reversal.

The RG flow for anisotropic superconductors has been first discussed by
Weinberg~\cite{Weinberg} at zero temperature, whereas the RG analysis turns out
to be more effective around the critical temperature $T_c$. An extension
to finite temperature of Shankar's~\cite{Shankar} derivation of the flow
equations is quite straightforward, and we shall focus on the main
aspects, thus referring to Shankar's pedagogical review~\cite{Shankar} for the
details.

The partition function $Z$ of a many-fermion system can be expressed as
the functional integral:
\begin{equation}
Z= \int \D \bar{\psi} \D \psi\, \e^{S[\psi,\bar{\psi}]} ,
\end{equation}
being $S = S_0 + S_{\mathrm int}$
the action as a functional of the spinor Grassmann fields
\begin{equation}
\psi\equiv
\left(\begin{array}{c}
\psi_\uparrow (\bk,\omega_n )\\
\psi_\downarrow (\bk,\omega_n )
\end{array}\right),~~~~\bar{\psi} \equiv
\left(\begin{array}{cc}
\bar{\psi}_\uparrow (\bk,\omega_n ) &
\bar{\psi}_\downarrow (\bk,\omega_n )
\end{array}\right),
\end{equation}
being $\bk$ a reciprocal lattice vector, $\omega_n = (2n+1)\pi/\beta$
a fermionic Matsubara frequency, $\beta=T^{-1}$ the inverse temperature
(hereafter, we set $\hbar = k_{\mathrm B} =1$).
The part of the action which is quadratic in the fields accounts for the
free evolution of the system, and may be written as:
\begin{equation}
\label{eq:freeaction}
S_0 = \frac{1}{\beta} \sum_\sigma \sum_n
\int\frac{\d^3 \bk}{(2\pi )^3}
\bar{\psi}_\sigma (\bk,\omega_n ) [i\omega_n +\mu -\varepsilon (\bk)]
\psi_\sigma (\bk, \omega_n ) ,
\end{equation}
being $\ve(\bk)$ the free single-particle dispersion relation.

The most general two-particle interaction in $d=3$ dimensions
is seen to contribute to the action through the term:
\begin{equation}
\label{eq:interactionaction}
S_{\mathrm int} = \frac{1}{2!2!} \int
\d(4)\d(3)\d(2)\d(1) \tilde{\delta} (4321)
\bar{\psi}(4) \bar{\psi}(3)
u(4321) \psi(2) \psi(1) ,
\end{equation}
being
\begin{equation}
\int \d(i) \equiv \frac{1}{\beta}
\sum_{\sigma_i} \sum_{n_i} \int \frac{\d^3 \bk_i}{(2\pi)^3} ,~~~~
\psi(i) \equiv \psi_{\sigma_i} (\bk_i ,\omega_{n_i} )
\end{equation}
and
\begin{equation}
\tilde{\delta} (4321) \equiv
(2\pi )^3 \beta \tilde{\delta}^{(3)} (\bk_4 +\bk_3 -\bk_2 -\bk_1 )
\delta_{n_4 + n_3 ,n_2 +n_1} ,
\end{equation}
where the Dirac $\tilde{\delta}^{(3)}$ enforces momentum conservation
up to a vector in the reciprocal lattice.
The interaction $u(4321)\equiv u_{\sigma_4 \sigma_3 \sigma_2 \sigma_1}
(\bk_4 ,\bk_3 ,\bk_2 ,\bk_1 )$ satisfies the general symmetry properties:
\begin{equation}
\label{eq:usymmetry}
u(4321)=-u(3421)=-u(4312)=u(3412)
\end{equation}
and acts as a $2\times 2\times 2\times 2$ array in the spinor
space.

Central to the RG approach is the assumption that all the integrals
may be evaluated within the shell
$|\ve(\bk)-\mu|<\Lambda$, with $\Lambda$ a proper energy cut-off,
provided all the parameters in the model are renormalized, in order to take
into account the elimination of the modes with $|\ve(\bk)-\mu|>\Lambda$.
Thus we suppose that all such parameters are renormalized to the scale
$\Lambda$
from the beginning. Of course, a ``physical'' choice of the cut-off $\Lambda$
would allow for a comparison of the parameters with the ``real world''
corresponding quantities.
A cut-off in energy is preferred, instead of a direct cut-off
in momenta, when dealing with a rotationally non-invariant Fermi surface.

When a low-temperature fermionic system is considered, the energy cut-off
$\Lambda$ can be chosen small enough to allow a unique decomposition for any
vector $\bk$ belonging to the tiny slice around the Fermi surface
$|\ve(\bk)-\mu|<\Lambda$ as:
\begin{equation}
\bk = \bk_0 + \delta\bk ,
\end{equation}
being $\bk_0$ the vector belonging to the Fermi surface ($\ve(\bk_0 ) -\mu =0$)
nearest to $\bk$ and $\delta\bk$ a vector orthogonal to the Fermi surface.
Expanding $\ve(\bk)$ around $\bk_0$, only the linear term shows to be
non-irrelevant in the RG flow, and one can safely write:
\begin{equation}
\ve(\bk)\approx \mu + v_{\mathrm F} (\bk_0 ) \delta k ,
\end{equation}
being $v_{\mathrm F} (\bk_0 ) = |\nabla_\bk \ve |_{\bk=\bk_0}$ the Fermi
velocity. The cut-off condition is then equivalently written as:
\begin{equation}
|v_{\mathrm F} (\bk_0 )\delta k |<\Lambda .
\end{equation}
We may always choose a bidimensional `vector' of parameters $\bth\equiv
(\theta_1 ,\theta_2 )$ on the Fermi surface to individuate $\bk_0$, together
with the energy displacement from the Fermi surface $\ve=\ve(\bk)-\mu =
v_{\mathrm F} (\bk_0 )\delta k$ to fix $\delta k$. Such a change of variables
is taken into account within the integrations according to the rule:
\begin{equation}
\int \frac{\d^3 \bk}{(2\pi)^3} \mapsto
\int \frac{\d^2 \bth}{(2\pi)^2} \int_{-\Lambda}^{\Lambda}
\frac{\d\ve}{2\pi} J(\bth,\ve),
\end{equation}
where
\begin{equation}
\label{eq:Jacobian}
J^{-1} (\bth,\ve) = \left|
\frac{\partial (\bth,\ve)}{\partial (\bk)}
\right| = \left| \nabla_\bk \ve \cdot(
\nabla_\bk \theta_1 \times \nabla_\bk \theta_2 )\right|
\end{equation}
is the inverse Jacobian function. In practice, we shall omit the $\ve$
dependence in $J$, since it reveals itself irrelevant in an RG sense,
by setting $J(\bth)\equiv J(\bth,\ve=0)$, which is the only marginal
term in the expansion of $J$ in powers of $\ve$.

The free action Eq.~(\ref{eq:freeaction}) now reads:
\begin{equation}
S_0 =
\frac{1}{\beta} \sum_{n,\sigma}
\int\frac{\d^2 \bth}{(2\pi )^2} J(\bth)
\int_{-\Lambda}^{\Lambda} \frac{\d\ve}{2\pi}
\bar{\psi}_\sigma (\bth,\ve,\omega_n )
[i\omega_n -\varepsilon]
\psi_\sigma (\bth,\ve, \omega_n ) .
\end{equation}

An elementary step in the RG flow is next defined as that transformation on
$S_0$ which: \emph{i)} integrates all the modes having $|\ve|<\Lambda /s$,
with $s\gtrsim 1$; \emph{ii)} rescales energies and momenta
as $s\omega_n \mapsto \omega^\prime_n$, $s\ve\mapsto \ve^\prime$, in order to
restore the original cut-off; \emph{iii)} rescales the fields as
$s^{-3/2} \psi\mapsto\psi^\prime$. At $T=0$, such a transformation leaves
the free action invariant, so that the latter can be considered as a fixed
point.

At a finite temperature, rescaling the energies as $s\omega_n \mapsto
\omega^\prime_n$ implies a rescaling of the inverse temperature itself
as $\beta/s \mapsto \beta^\prime$. This is what we must pay for
having restored the cut-off to its original value. Besides, if any energy
scale is associated with the system, such as a non-zero temperature or an
energy gap, this one undergoes a renormalization flow, while the cut-off
$\Lambda$ is kept fixed. If one identifies $s=\e^t$, then:
\begin{equation}
\label{eq:betaflow}
\frac{\d\beta_t }{\d t} = -\beta_t ~~\Leftrightarrow~~
 \beta_t =\beta_0 \e^{-t} ,
\end{equation}
which shows that renormalizing the action at a \emph{fixed}
energy cut-off $\Lambda$ yields a \emph{larger} effective temperature.

A completely equivalent approach consists in keeping fixed all
the energy scales (temperature, gap {\em etc}), while assuming a flow
in the cut-off as:
\begin{equation}
\Lambda_t =\Lambda_0 \e^{-t} .
\end{equation}
One approach is recovered from the other by  a bare
change of variables in the integration over $\ve$.

Let us now examine how such a RG transformation affects the interaction
part of the action functional, Eq.~(\ref{eq:interactionaction}).
At a tree level (Fig.~\ref{fig:treelevel}), the only marginal couplings which
survive the RG flow are the lowest order terms in the expansion of
$u(4321)$ in powers of $\ve_i$. Moreover, from momentum conservation
and phase space arguments, they restrict to only two contributions:
\begin{eqnarray}
F(\bth_1 ,\bth_2 ) && =
u(4321)~\mbox{with}~\ve_i =0,~\hat{\Omega}_4 \cdot \hat{\Omega}_3 =
\hat{\Omega}_2 \cdot \hat{\Omega}_1 ,
{}~~~~~~~~~~~~~\mbox{($F$ coupling),} \\
V(\bth_1 ,\bth_4 ) && =
u(4321)~\mbox{with}~\ve_i =0,~\hat{\Omega}_4 + \hat{\Omega}_3 =
\hat{\Omega}_2 + \hat{\Omega}_1 =0 ,
{}~~~~\mbox{($V$ coupling),}
\end{eqnarray}
where $\hat{\Omega}_i$ denotes the unit vector in the direction of
$\bk_i \equiv (\bth_i , \ve_i =0)$.

We are interested in the flow of the $V$-couplings which are responsible
for the superconductive instability of the Fermi surface. We assume the
interaction to be spin independent, and we take in consideration only spin
singlet pairs, so that we can get rid of all the spin indices, by posing:
\begin{eqnarray}
V(\bth_1 ,\bth_4 ) \equiv V(\bth_4 ,\bth_1 ) &&=
\frac{1}{2} [
u_{\uparrow\downarrow\downarrow\uparrow}
(\bk_4 ,-\bk_4 , -\bk_1 ,\bk_1 )
-
u_{\uparrow\downarrow\uparrow\downarrow}
(\bk_4 ,-\bk_4 , -\bk_1 ,\bk_1 )] =\nonumber\\
&& =
\frac{1}{2} [
u_{\downarrow\uparrow\uparrow\downarrow}
(\bk_4 ,-\bk_4 , -\bk_1 ,\bk_1 )
-
u_{\downarrow\uparrow\downarrow\uparrow}
(\bk_4 ,-\bk_4 , -\bk_1 ,\bk_1 )] ,
\label{eq:potentialdefinition}
\end{eqnarray}
being the $u$ interactions restricted on the Fermi surface. Besides,
the general symmetry properties, Eq.~(\ref{eq:usymmetry}), imply
that $V(\bth_1 ,\bth_4 )$ is symmetric with respect to the inversion
$\hat{\Omega}_1 \rightarrow - \hat{\Omega}_1$ or $\hat{\Omega}_4
\rightarrow -\hat{\Omega}_4$.

Following analogous arguments as for $T=0$,~\cite{Shankar} we observe that
at one-loop level only the {\sc bcs} diagram contributes to the flow in $V$
(Fig. \ref{fig:onelooplevel}) with:
\begin{equation}
\label{eq:oneloopcorrection}
\d V(\bth_1 ,\bth_4 ) = -\frac{1}{2}
\int_{\d\Lambda}  \d(6)\d(5)
G^0 (6) G^0 (5)
V(\bth_1 ,\bth_5 ) V(\bth_5 ,\bth_4)
\tilde{\delta} (6521) ,
\end{equation}
where $G^0 (i)$ denotes the free fermion propagator, which at a finite
temperature may be written as:
\begin{equation}
\label{eq:propagator}
G^0 (i) \equiv G^0 (\bk_i ,\omega_{n_i} ) =
\frac{1}{i\omega_{n_i} -\ve (\bk_i ) +\mu} ,
\end{equation}
and where $\int_{\d\Lambda} \d(6)\d(5)$ does not contain the sum over
spin indices, and the integration is performed
over the tiny slice $|\ve(\bk) \pm \Lambda|< |\d\Lambda |$.

For a generic anisotropic Fermi surface, when an interaction is switched on,
both the chemical potential and the shape of the Fermi surface change according
to the definition:~\cite{Luttinger}
\begin{equation}
\mu^\prime -\ve(\bk ) - \mathop{\mathrm Re} \Sigma (\bk,\mu^\prime ) =0,
\end{equation}
being $\mu^\prime$ the chemical potential in presence of the interaction and
$\Sigma(\bk,\omega)$ the self-energy in the interacting fermion propagator.
At one-loop level, a contribution to $\Sigma$ comes from the `tadpole' diagram
shown in Fig.~\ref{fig:treelevel}. However, the propagator in the one-loop
correction Eq.~(\ref{eq:oneloopcorrection}) is a free propagator, and its
denominator in Eq.~(\ref{eq:propagator}) is determined in terms of the free
single-particle dispersion relation and chemical potential. So, no serious
problem arises at one-loop from the renormalization of the Fermi surface.

In Eq.~(\ref{eq:oneloopcorrection}), the incoming momentum and frequency are
zero, and therefore the loop momenta and frequencies are opposite. From time
reversal, $\ve(\bk)=\ve(-\bk)$, and taking into account the $\tilde{\delta}$
constraint, one gets:
\begin{eqnarray}
\d V(\bth_1 ,\bth_4 ) && =
-\frac{1}{\beta} \sum_n \int \frac{\d^2 \bth}{(2\pi)^3}
J(\bth) V(\bth_1 ,\bth) V(\bth,\bth_4 )
\int_{\Lambda-\d\Lambda}^\Lambda \frac{\d\ve}{\ve^2 +\omega_n^2} =\nonumber\\
&& = -\int \frac{\d^2 \bth}{(2\pi)^3} J(\bth) V(\bth_1 ,\bth)V(\bth,\bth_4 )
\int_{\Lambda-\d\Lambda}^\Lambda \frac{\d\ve}{2\ve} \tanh\frac{\beta\ve}{2} ,
\end{eqnarray}
where we made use of the Mittag-Leffler expansion:~\cite{Gradshtein}
\begin{equation}
\frac{1}{\beta} \sum_n \frac{1}{\omega_n^2 + \ve^2} =
\frac{1}{2\ve} \tanh \frac{\beta\ve}{2} .
\end{equation}

Integrating at the cut-off, with $\d t= |\d\Lambda |/\Lambda$, one finally
obtains:
\begin{equation}
\label{eq:Vflow}
\frac{\d V_t (\bth_1 ,\bth_2 )}{\d t} =
-\frac{1}{2}
\tanh\left(\frac{\Lambda\beta_t}{2} \right)
\int \frac{\d^2 \bth}{(2\pi)^3}  J(\bth)
V_t (\bth_1 ,\bth) V_t (\bth,\bth_2 ) ,
\end{equation}
where $\beta_t$ flows according to Eq.~(\ref{eq:betaflow}).

We notice that the choice of a cut-off in energy allows one to take
the factor $\tanh (\Lambda\beta_t /2)$ out of the integration over the Fermi
surface, like in the rotationally invariant case. For $\beta_0 \rightarrow
\infty$, Eq.~(\ref{eq:Vflow}) reduces to the flow equation for an anisotropic
superconductor recovered by Weinberg~\cite{Weinberg} at $T=0$. Besides,
$T$ increases monotonically as $t$ flows down to $\infty$, so that
$T\rightarrow\infty$ ($\beta=0$) is always a fixed point, since
Eq.~(\ref{eq:Vflow})
would then yield $\d V_t /\d t =0$.

Introducing the measure
\begin{equation}
\d\tau_\bth \equiv \frac{\d^2 \bth}{2(2\pi)^3} J(\bth),
\end{equation}
one may regard Eq.~(\ref{eq:Vflow}) as the flow equation for the integral
operator:
\begin{equation}
\left( \hat{V}_t \Phi \right) (\bth) = \int\d\tau_{\bth^\prime}
V_t (\bth,\bth^\prime ) \Phi (\bth^\prime ),
\end{equation}
where the coupling $V_t (\bth,\bth^\prime )$ plays the r\^ole of a symmetric
Hilbert-Schmidt kernel.~\cite{Kolmogorov} Eq.~(\ref{eq:Vflow}) then reads:
\begin{equation}
\label{eq:Voperatorflow}
\frac{\d \hat{V}_t}{\d t} = - \tanh \left(\frac{\Lambda\beta_t}{2} \right)
\hat{V}_t \cdot \hat{V}_t .
\end{equation}
The integral operator $\hat{V}_t$ is Hermitean and admits a complete set of
orthonormal eigenfunctions $\Phi_\alpha$:
\begin{eqnarray}
\label{eq:eigenvalue}
\left( \hat{V}_t \Phi_\alpha \right) (\bth) && = \lambda_t (\alpha)
\Phi_\alpha (\bth ) \\
\label{eq:orthonormality}
\int\d\tau_\bth \Phi_\alpha^* (\bth) \Phi_\beta (\bth) && =
\delta_{\alpha\beta} .
\end{eqnarray}

As suggested by Weinberg,~\cite{Weinberg} Eq.~(\ref{eq:Voperatorflow}) implies
that $[\d \hat{V}_t ,\hat{V}_t ] =0$, so that the eigenfunctions $\Phi_\alpha
(\bth)$ do not flow, whereas the eigenvalues do flow according to the
(decoupled) equations:
\begin{equation}
\label{eq:lambdaflow}
\frac{\d\lambda_t (\alpha)}{\d t} = -\tanh\left( \frac{\Lambda\beta_t}{2}
\right) \lambda^2_t (\alpha),
\end{equation}
which can be integrated together with Eq.~(\ref{eq:betaflow}) to yield:
\begin{equation}
\label{eq:lambdat}
\lambda_t = \frac{\lambda_0}{1 + \lambda_0
\int_0^t \tanh\left( \frac{\Lambda\beta_0}{2} \e^{-\tau} \right) \d\tau} ,
\end{equation}
being $\lambda_0$ the unrenormalized eigenvalue. The fixed point
at $\beta=0$ is generally reached unless $\lambda_t$ diverges for some
$\bar{t}$, where the flow stops. This may only occur if the unrenormalized
eigenvalue were negative at the beginning of the flow, $\lambda_0 <0$.
As shown in Fig.~\ref{fig:flow}, at variance with the $T=0$ analysis,
this instability may not occur if $\lambda_0$ is small, and
a suitable definition of the critical temperature $T_c$ can therefore
be obtained from Eq.~(\ref{eq:lambdat}) in the limit
$\bar{t}\rightarrow\infty$,
\emph{i.e.,}
\begin{equation}
1 + \lambda_0
\int_0^\infty \tanh\left( \frac{\Lambda\beta_0}{2} \e^{-\tau} \right) \d\tau =0
,
\end{equation}
or, after a change of variables,
\begin{equation}
\label{eq:lambdaxi}
\frac{1}{\lambda_0} = - \int_0^{\Lambda\beta_c}
\frac{\d\xi}{\xi} \tanh \frac{\xi}{2} ,
\end{equation}
with $\lambda_0$ the negative eigenvalues being largest in modulus,
\emph{i.e.} the first leading to a divergence in $\lambda_t$.

Eq.~(\ref{eq:lambdaxi}) is quite familiar, and in the limit
$\Lambda\beta_c \ll 1$ may be analytically approximated as:
\begin{equation}
\label{eq:Tc}
T_c = \frac{2\e^\gamma}{\pi} \Lambda \e^{1/\lambda_0} ,
\end{equation}
being $\gamma\simeq 0.5772$ the Catalan-Euler's constant.

We recover the {\sc bcs} limit for a constant coupling $V=-2v$, since
then:
\begin{equation}
\lambda_0 = -2v \int \d\tau_\bth = -nv,
\end{equation}
being $n$ the density of states per spin at the Fermi energy. In this
limit, a natural cut-off is provided by the Debye frequency. Moreover,
Eq.~(\ref{eq:lambdaxi}) is very general, and holds for any anisotropic
superconductor. The critical temperature is determined by the most
negative eigenvalue of the coupling kernel $V_t (\bth,\bth^\prime )$, while
the remaining eigenvalues do not play any r\^ole in the proximity of the
critical point. This idea will be developed in the next section, where
the r\^ole of the eigenfunctions will be clarified in connection with
the scale invariant gap equation.

\section{Mean field \emph{vs} RG: a scale invariant gap equation}

In the mean-field theory of superconductivity, one is usually
led to consider the energy gap function $\Delta(\bk)$.
It is a quantity of central interest, since it is directly
measurable and its symmetry patterns are
strongly related to the nature of the pairing, as we shall see in
the following.

It would be desirable that the RG analysis could give some indication
about the nature and the size of the gap function. On the other hand, we
would like to throw some light on the connection between a RG approach
and a conventional mean-field approximation.

At the end of the previous Section, we noticed that only the most negative
eigenvalue of the integral operator $\hat{V}_t$ is responsible for the
location of the instability. We are therefore led to expect that the
associated eigenfunction should determine the gap structure at the transition
point. Indeed, expanding the integral operator $\hat{V}_t$
in terms of its orthonormalized eigefunctions:
\begin{equation}
\label{eq:expansion1}
\left( \hat{V}_t ~\cdot~ \right) (\bth) =
\sum_\alpha \Phi^*_\alpha (\bth) \lambda_t (\alpha)
\int \d\tau_{\bth^\prime} \Phi_\alpha (\bth^\prime ) ~\cdot~ ,
\end{equation}
we may decompose the coupling kernel as $V_t (\bth,\bth^\prime ) =
V_t^c (\bth,\bth^\prime ) + \tilde{V}_t (\bth,\bth^\prime )$, where:
\begin{eqnarray}
V_t^c (\bth,\bth^\prime ) && = \Phi_0^* (\bth) \lambda_t (0) \Phi_0
(\bth^\prime ) \nonumber\\
\label{eq:expansion}
\tilde{V}_t (\bth,\bth^\prime ) &&= \sum_{\alpha\neq 0}
\Phi_\alpha^* (\bth) \lambda_t (\alpha) \Phi_\alpha (\bth^\prime ) ,
\end{eqnarray}
being $\lambda_t (0)$ the most negative eigenvalue of $\hat{V}_t$.
The divergence of $\lambda_t (0)$ at the transition point as
$t\rightarrow\infty$ clearly indicates the relative importance of $V_t^c$
in driving the instability in the proximity of the transition point.
In other words, at the transition point, the kernel $V_t (\bth,\bth^\prime )$
may be approximated with its `critical' part $V_t^c (\bth,\bth^\prime )$,
whose structure is entirely determined by the eigenfunction $\Phi_0 (\bth)$
belonging to $\lambda_t (0)$.

The very same conclusion may be reached through a completely different
path, starting
from the standard mean-field gap equation:~\cite{Fetter}
\begin{eqnarray}
\label{eq:gapequation}
\Delta (\bk) &&= -\frac{1}{2} \int \frac{\d^3 \bk}{2(2\pi)^3}
V(\bk,\bk^\prime ) \frac{\Delta(\bk^\prime )}{%
E(\bk^\prime )} \tanh\frac{\beta E(\bk^\prime )}{2},\\
E (\bk)&&=
\sqrt{\Delta^2 (\bk)+ \left[ \ve(\bk) -\mu \right]^2} ,\nonumber
\end{eqnarray}
where $V(\bk,\bk^\prime )$ is
understood as an extension of the kernel $V(\bth,\bth^\prime )$ out of
the Fermi surface. If $\Delta$ is small, compared to the other proper
energies of the system, we may assume that the important contribution to
the integral Eq.~(\ref{eq:gapequation}) comes from the region around
the Fermi surface. Therefore, introducing a cut-off $\Lambda\gg\Delta$,
substituting the renormalized $\hat{V}_t$ kernel for the unrenormalized
interaction, and neglecting the energy dependence of $\Delta$ and $V$,
we may eventually write Eq.~(\ref{eq:gapequation}) as:
\begin{equation}
\label{eq:gap}
\Delta_t (\bth) = - \frac{1}{2}\int \d\tau_{\bth^\prime}
V_t (\bth,\bth^\prime ) \Delta_t (\bth^\prime ) \int_{-\Lambda}^{\Lambda}
\d\ve \frac{1}{\sqrt{\ve^2 + \Delta_t^2 (\bth^\prime )}}
\tanh \left[\frac{\beta}{2} \sqrt{\ve^2 + \Delta_t^2 (\bth^\prime )} \right].
\end{equation}

An explicit $t$ dependence had to be attached to $\Delta$ and $\beta$,
albeit  trivial, being a consequence of choosing to keep $\Lambda$
fixed during the flow: both $\Delta$ and $\beta^{-1}$ are energy scales
of the system, and are therefore expanded as $\Delta_t = \Delta\e^t$,
$\beta_t^{-1} = \beta^{-1} \e^t$ during the flow. In the present context,
it would be preferable to work with a flowing cut-off $\Lambda_t =
\Lambda_0 \e^{-t}$ and fixed values of the energy scales $\Delta$, $\beta$,
so that changing variables $\ve\mapsto\ve\e^t$ we may write Eq.~(\ref{eq:gap})
as:
\begin{equation}
\label{eq:gapfixed}
\Delta(\bth) = -\frac{1}{2} \int\d\tau_{\bth^\prime}
V_t (\bth,\bth^\prime ) \Delta (\bth^\prime )
\int_{-\Lambda_t}^{\Lambda_t} \d\ve
\frac{1}{\sqrt{\ve^2 +\Delta^2 (\bth^\prime )}}
\tanh\left[\frac{\beta}{2} \sqrt{\ve^2 +\Delta^2
(\bth^\prime )}\right].
\end{equation}

We first discuss Eq.~(\ref{eq:gapfixed}) near the transition point,
$\beta\rightarrow\beta_c$,
where the RG approach shows itself more effective than at $\beta\gg\beta_c$.
In a mean-field framework,
the critical point would then be defined just by the vanishing of the gap
function. Without loss of generality, we may pose:
\begin{equation}
\label{eq:chi}
\Delta(\bth)=\Delta_c \chi(\bth),
\end{equation}
being $\chi(\bth)$ regular at the critical point and $\Delta_c$ a scale
parameter, $\Delta_c = {\mathcal O}(\beta- \beta_c )$. Insertion of
Eq.~(\ref{eq:chi}) into Eq.~(\ref{eq:gapfixed}) at the critical point
yields:
\begin{equation}
\chi(\bth) = -\int_0^{\Lambda_t} \frac{\d\ve}{\ve}
\tanh \left( \frac{\beta_c \ve}{2} \right)
\int \d\tau_{\bth^\prime} V_t (\bth,\bth^\prime ) \chi (\bth^\prime ) ,
\end{equation}
which shows that $\chi(\bth)$ is the eigenfunction of the integral operator
$\hat{V}_t$ belonging to the eigenvalue $\lambda_t$ such that:
\begin{equation}
\label{eq:lambdaxi2}
\frac{1}{\lambda_t} = -\int_0^{\Lambda_t \beta_c} \frac{\d\xi}{\xi}
\tanh \frac{\xi}{2} .
\end{equation}
The latter result makes sense only if $\lambda_t$ is the most negative
eigenvalue of $\hat{V}_t$, in which case Eq.~(\ref{eq:lambdaxi2}) is identical
to Eq.~(\ref{eq:lambdaxi}), thus yielding the same characterization for
the critical temperature, Eq.~(\ref{eq:Tc}).

As we expected, the functional form of the gap function near the critical point
is fixed as:
\begin{equation}
\Delta(\bth) =\Delta_c \Phi_0 (\bth).
\end{equation}
We notice that substituting the critical part $V_t^c$ for $V_t$ in
the gap equation (\ref{eq:gapfixed}) yields the same result for both
the critical temperature and the gap structure. The latter finding
is of particular interest when the eigenfunctions display different
symmetries, since only one of them is seen to rule
over the symmetry pattern of the energy gap, so that no wave mixing,
such as $s$-$d$, may occur as $T\rightarrow T_c$.

We remark that Eq.~(\ref{eq:lambdaxi2}) stems from a gap
equation linearization near the critical point, so that
its agreement with Eq.~(\ref{eq:lambdaxi}) points forward
a substantial equivalence of the RG apporach with the
mean-field approximation. However, the former approach
displays the remarkable advantage of characterizing
the transition in terms of the properties of the system
near the Fermi surface. Namely, the shape of the latter
together with the interaction kernel evaluated at the
Fermi energy entirely determine the RG flow. In this
sense, the RG approach allows to classify within a
unified scheme the instabilities of fermionic systems
sharing the same behavious near the Fermi surface,
although different in nature.

On the other hand, integrating modes far from the Fermi
surface provides a simplified description of the system
which still contains all its relevant features. Besides,
this description will prove itself handier when dealing
with the numerical implementations, as in the next Section.

In general, inserting the expansion Eq.~(\ref{eq:expansion1})  in the
gap equation (\ref{eq:gapfixed}), we may formally write:
\begin{equation}
\label{eq:gapexpanded}
\Delta(\bth) = -\frac{1}{2} \sum_\alpha \Phi^*_\alpha (\bth) \lambda_t (\alpha)
\int\d\tau_{\bth^\prime}
\Phi_\alpha (\bth^\prime )
\Delta(\bth^\prime ) \int_{-\Lambda_t}^{\Lambda_t}
\d\ve \frac{1}{\sqrt{\ve^2 +\Delta^2 (\bth^\prime )}}
\tanh\left[\frac{\beta}{2}\sqrt{\ve^2 +\Delta^2 (\bth^\prime )}
\right],
\end{equation}
which means that, at any $T<T_c$, the gap function is a linear combination
of eigenfunctions of $\hat{V}_t$. This is not a trivial statement, since
the eigenfunctions appearing in the expansion belong to non-zero
eigenvalues, whose number may be finite in several physically relevant cases,
\emph{i.e.} when the kernel is separable.

Eq.~(\ref{eq:gapexpanded}) is highly non-linear, and may possess more
than one solution. For instance, broken-symmetry solutions may occur
as linear combination of eigenfunctions belonging to different invariant
subspaces. However, in the limit $\beta\rightarrow\beta_c$, the physical
solution tends to a specific eigenfunction with a fixed symmetry,
thus preventing the occurrence of a broken-symmetry gap function at the
critical point.

Far from the transition point, the present analysis does not add too much
to the general comprehension of the problem. For instance, at zero temperature
the internal integral can be evaluated in Eq.~(\ref{eq:gapexpanded}) thus
yielding, in the limit $\Delta\ll\Lambda$:
\begin{equation}
\label{eq:gapformal}
\Delta(\bth) = -\sum_\alpha \Phi_\alpha (\bth) \lambda_t (\alpha)
\int\d\tau_{\bth^\prime} \Phi_\alpha (\bth^\prime )\Delta (\bth^\prime )
\log \left| \frac{2\Lambda_t}{\Delta(\bth^\prime )} \right|.
\end{equation}

The latter gap equation, valid in the limit $\Delta\ll\Lambda$, can be easily
proven to be scale invariant according to the flow equation
(\ref{eq:lambdaflow}),
and is completely equivalent to the scale invariant gap equation
recovered by Weinberg.~\cite{Weinberg} Moreover, we write the expansion
Eq.~(\ref{eq:gapexpanded}) as:
\begin{equation}
\label{eq:gapexpanded2}
\Delta(\bth) = \sum_\alpha \Phi_\alpha^* (\bth)\Delta(\alpha),
\end{equation}
whose insertion into
Eq.~(\ref{eq:gapformal}), making use of the linear
independence of the eigenfunctions, yields:
\begin{equation}
\label{eq:coupled}
\Delta(\alpha) =-\lambda_t (\alpha) \sum_{\alpha^\prime} \Delta(\alpha^\prime )
\langle\alpha^\prime
|\log\left|\frac{2\Lambda_t}{\Delta(\bth)}\right| |\alpha\rangle,
\end{equation}
being $\langle\alpha^\prime |f(\bth)|\alpha\rangle \equiv \int\d\tau_\bth
\Phi_{\alpha^\prime}^* (\bth) f(\bth) \Phi_\alpha (\bth)$, which is a set of
non-linear coupled equations for the coefficients $\Delta(\alpha)$.
Multiplying times $\Delta^* (\alpha)/\lambda_t (\alpha)$, summing over
$\alpha$, and introducing the scale $\Delta_0$ defined as:~\cite{Weinberg}
\begin{equation}
\label{eq:delta01}
\int\d\tau_\bth |\Delta(\bth)|^2 \log\left|\frac{\Delta_0}{\Delta(\bth)}
\right| =0,
\end{equation}
one eventually obtains:
\begin{equation}
\label{eq:delta02}
\Delta_0 = 2\Lambda_t \exp \left(
\frac{\sum_\alpha |\Delta(\alpha)|^2 /\lambda_t (\alpha)}{%
\sum_\alpha |\Delta(\alpha)|^2 }
\right),
\end{equation}
which is valid only in the limit $\Delta_0 \ll \Lambda_t$. We observe
that far from the critical point all the components of the coupling
kernel $V_t$ contribute to the determination of the gap function.
In practice, the present approach has the advantage of dealing with a
set of coefficients $\Delta(\alpha)$ whose number may be finite, in the
case of a separable kernel. However, we cannot escape from a full solution
of the set of the non-linear coupled equations (\ref{eq:coupled}).
The present decomposition in eigenfunctions is the generalization of the
standard decomposition in eigenfunctions of the angular momentum, which
are simultaneous eigenfunctions of $\hat{V}_t$ only for a fully
rotationally invariant system.

If some symmetry invariance is present in both the interaction and in
the Fermi surface, then the set of equations (\ref{eq:coupled}) may be
partially decoupled: in other words, if $\{\Phi_\alpha (\bth)\}$ is
an invariant subspace, and $\hat{U}_g$ is a unitary representation of
the symmetry group $\{ g\}$, \emph{i.e.} $[\hat{U}_g ,\hat{V}_t ]=0$
and $\{\hat{U} (g)\Phi_\alpha (\bth)\} \equiv\{\Phi_\alpha (\bth)\}$,
$\forall g\in\{ g\}$, then we may require that the gap function shares
the same symmetry:
\begin{equation}
\hat{U} (g) \Delta(\bth) = \e^{i\varphi(g)} \Delta(\bth) ,~~~
\forall g\in\{ g\} ~~~\mbox{($|\Delta(\bth)|^2$ invariant),}
\end{equation}
which is equivalent to saying that:
\begin{equation}
\hat{U} (g) \Phi_\alpha (\bth) = \e^{i\varphi(g)} \Phi_\alpha (\bth),~~~
\forall\Phi_\alpha  \in\{\Phi_\alpha \}
\end{equation}
(the eigenfunctions $\{\Phi_\alpha \}$ belong to the same eigenvalue
$\e^{i\varphi(g)}$) and $\Delta(\bth) =\sum_\alpha \Delta(\alpha)
\Phi_\alpha^* (\bth)$, with $\Phi_\alpha$ belonging to the invariant
subset $\{\Phi_\alpha \}$.

The set of equations (\ref{eq:coupled}) does admit such solutions with
a given symmetry (eigenfunctions of $\hat{U}$) since the matrix elements
$\langle\alpha^\prime |\log|2\Lambda_t /\Delta ||\alpha\rangle$ vanish if the
functions $\Phi_\alpha$, $\Phi_{\alpha^\prime}$
belong to different invariant subspaces,
provided that $|\Delta|$ is invariant.

Of course the symmetry invariance of the integral operator $\hat{V}_t$ does
not prevent from the occurrence of broken-symmetry solutions, since the
non-linearity of the set of equations (\ref{eq:coupled}) does not guarantee
the uniqueness of the symmetric solution.

\section{A tight-binding model for layered superconductors}

The anisotropic flow equations, derived at a finite temperature in the
previous Sections, are here employed in the framework of a simple
tight-binding model recently proposed for describing the band
structure and superconductive properties of
Bi$_2$Sr$_2$CaCu$_2$O$_{8+\delta}$
(BSCCO).~\cite{Schneider,Spathis}
Alike the majority of the high $T_c$ cuprate superconductors, BSCCO
is characterized by a layered structure, which gives rise to
many anisotropic
physical properties, and further results in a rotationally non-invariant
hole dispersion relation $\ve(\bk)$.

Allowing for nearest-neighbour inter-plane
hopping and for nearest-neighbour and next-nearest-neighbour intra-plane
hopping, in the usual tight-binding approximation, $\ve(\bk)$ reads
as:~\cite{Schneider,Spathis}
\begin{equation}
\label{eq:dispersionrelation}
\ve (\bk) = A\left[
-2(\cos k_x + \cos k_y ) +4B \cos k_x \cos k_y
-2C \cos k_z \right] ,
\end{equation}
where the components $k_x ,k_y ,k_z$
of the wave-vector $\bk$ are measured in units
of the respective inverse lattice spacings.
The constants $A,B,C$ have been determined by comparison with
photoemission data~\cite{Schneider,Spathis} to be
$A=0.05~e{\mathrm V}$, $B=0.45$, $C=0.1$
in order to reproduce the observed hole density of states for BSCCO.
In particular, the condition
$C\ll B\ll 1$ denotes a smaller intra-plane next-nearest-neighbour
hopping probability than a nearest-neighbour
one, and an even smaller inter-plane nearest-neighbour one, due to the large
plane separation. In the following, we
shall measure all the energies in units of $A$.

In order to set up the model, we must assume the existence of a pairing
interaction, and the short coherence length would suggest the relevance
of the short range part of the interaction. Thus, on very general grounds,
neglecting the long range contribution, we may expand any pairing interaction
as the sum of an on-site term and of contributions arising from
nearest neighbour sites, next-nearest neighbours and so on.

We do not address any question concerning the physical origin of the
pairing interaction, but once the free fermion dispersion relation
has been fixed on a phenomenological basis, we rather wish
to explore the main physical consequences arising from the choice
of a pairing interaction. In other words, we keep the dispersion
relation fixed, and change the interaction in order to observe the
effects on the superconductive instability.

As an illustration of the method developed before, we here retain
only two terms in the expansion of the interaction, namely the
on-site and the in-plane nearest-neighbour singlet pairing couplings.
The effects of only an in-plane nearest-neighbour pairing have been already
considered, together with an identical dispersion relation, in the
framework of mean-field approximation, by Spathis \emph{et al.}~\cite{Spathis}

The Fourier expansion of the pairing interaction $V(\bk,\bk^\prime )$
reads as:
\begin{equation}
\label{eq:potential}
V(\bk,\bk^\prime ) = u_0 + 2u_x \cos (k_x -k^\prime_x ) +
2u_y \cos (k_y -k^\prime_y ),
\end{equation}
being $u_0$ the on-site Hubbard $U$ and $u_x$, $u_y$ the
nearest-neighbour interactions in the $x$ and $y$ directions, respectively.
Enforcing the general symmetry properties Eq.~(\ref{eq:usymmetry})
on the interaction matrix, the singlet pairing interaction results
in the symmetrized part of Eq.~(\ref{eq:potential}) with respect
to the inversion $\bk^\prime \rightarrow -\bk^\prime$ or
$\bk \rightarrow -\bk$,
\begin{eqnarray}
V(\bk,\bk^\prime ) &&= u_0
+u_x [\cos(k_x - k^\prime_x ) + \cos(k_x +k_x^\prime )]
+u_y [\cos(k_y - k^\prime_y ) + \cos(k_y +k_y^\prime )]  =\nonumber\\
\label{eq:potentialbis}
&&= u_0 +2 u_x \cos k_x \cos k_x^\prime
+2 u_y \cos k_y \cos k_y^\prime .
\end{eqnarray}

The Fermi surface,
defined as the locus of the points $\bk$ in the momentum space
verifying the equation $\ve(\bk) =\mu$, is a quite complicated,
definitely rotationally non-invariant surface, whose shape varies
furthermore with $\mu$. For
$\mu_1 = -4+4B-2C \le\mu\le-4+4B+2C =\mu_2$ it is a closed
surface, which for small values of the chemical potential $\mu$
can be approximated by an ellipsoid of square semiaxes
$\bar{k}_x^2 =\bar{k}_y^2 = (\mu+\mu_1)/(2B-1)$,
$\bar{k}_z^2 = -(\mu+\mu_1 )/C$, which reduces to
a sphere since $C=1-2B$. The Fermi surface
becomes an open surface as $\mu$ increases up to
the full bandwidth, $\mu_3 =4+4B+2C$, where it reduces to the eight zone
corners. Fig.~\ref{fig:FS} shows the Fermi surface
within the positive octant of the
first Brillouin zone for various values of $\mu$.

It seems convenient to choose $\bth=(k_x ,k_y )\equiv(\theta_x ,\theta_y )$
as suitable coordinates
upon the Fermi surface, for each value of $\mu$. Due to its varying shape at
increasing $\mu$, coordinates $\bth$ will be affected by limitations,
depending on $\mu$, which we shall keep understood in the following,
when writing integrals over the Fermi surface. The Jacobian function
Eq.~(\ref{eq:Jacobian}) may then be straightforwardly worked out as:
\begin{equation}
J^{-1} (\bth,\ve)=
\sqrt{4C^2 -\left(
\mu+\ve+2(\cos\theta_x +\cos\theta_y ) -4B\cos\theta_x \cos\theta_y
\right)^2} .
\end{equation}

The hole dispersion relation Eq.~(\ref{eq:dispersionrelation}) also fixes the
density of states of the system, in its normal phase, which may be expressed
as:
\begin{equation}
\label{eq:densityofstates}
n(\ve+\mu)= \frac{1}{2\pi} \int \frac{\d^2 \bth}{(2\pi)^2}
J(\bth)
=2\int\d\tau_\bth ,
\end{equation}
A further integration from the band bottom $\mu=\mu_1$ yields
the total fraction of occupied states:
\begin{equation}
\label{eq:occupiedstates}
N(\mu) = \int_{\mu_1}^\mu \d\ve~ n(\ve),
\end{equation}
together with the `normalization' condition at the top of the band,
$N(\mu_3 )=1.$

The quantities $n(\mu)$ and $N(\mu)$ have been numerically evaluated,
as functions of the chemical potential $\mu$,
$-\mu_1 \le\mu\le \mu_3$ (Fig.~\ref{fig:statedensa}, \ref{fig:statedensb}).
In particular, $n(\mu)$ displays a pronounced, yet finite, maximum
for $\mu=\mu_2 =-2.0$. It may be regarded as a token of the system's
quasi-bidimensionality, due to its layered structure, and would have been a
true van Hove singularity, resulting in an infinite peak, if no
inter-plane hopping had been considered, however small ($C=0$).

For our purposes, the pairing potential Eq.~(\ref{eq:potentialbis})
is recognized as
the symmetrical kernel
$V(\bth,\bth^\prime )$ in Eq.~(\ref{eq:potentialdefinition}),
provided that $k_x ,k_y \mapsto \theta_x ,\theta_y$,
$k^\prime_x ,k^\prime_y \mapsto \theta^\prime_x ,
\theta^\prime_y$, and that $\bth,\bth^\prime$ are
suitably restricted on the Fermi surface.
This kernel may then be put into the `separate' form:
\begin{equation}
\label{eq:separate}
V(\bth,\bth^\prime ) = \sum_{i,j=0,x,y} \eta_{ij} U_i (\bth) U_j (\bth^\prime
),
\end{equation}
with:
\begin{equation}
\label{eq:matrix}
\eta=\left(
\begin{array}{ccc}
u_0 &0&0\\
0&2u_x &0\\
0&0&2u_y
\end{array}\right)
\end{equation}
and:
\begin{equation}
U(\bth) = \left(
\begin{array}{c}
1\\
\cos\theta_x\\
\cos\theta_y
\end{array}\right).
\end{equation}

Insertion of Eq.~(\ref{eq:separate}) into Eq.~(\ref{eq:eigenvalue})
for the eigenvalue problem straightforwardly yields:
\begin{equation}
\Phi_\alpha (\bth)= \sum_{i=0,x,y} \gamma_i (\alpha) U_i (\bth).
\end{equation}
Further substitution yields:
\begin{equation}
\sum_{i=0,x,y} U_i (\bth) \left[ \lambda(\alpha)\gamma_i (\alpha)
-\sum_{j,k=0,x,y} \eta_{ij} \gamma_k (\alpha)
\int \d\tau_\bth U_j (\bth) U_k (\bth) \right] =0,
\end{equation}
which, due to the linear independence of the functions $U_i$, yields:
\begin{equation}
\label{eq:eigsys}
\sum_{k=0,x,y}
\gamma_k (\alpha) \left[ \lambda(\alpha) \delta_{ik}
-\sum_{j=0,x,y} \eta_{ij} \langle U_j U_k \rangle \right] =0
\end{equation}
with:
\begin{equation}
\label{eq:eigsysint}
\langle U_j U_k \rangle = \int \d\tau_\bth
U_j (\bth)U_k (\bth),
\end{equation}
being, in particular,
$2\langle U_0^2 \rangle (\mu) = n(\mu)$ (Fig.~\ref{fig:statedensa}).

The linear homogeneous system Eq.~(\ref{eq:eigsys}) allows then for non trivial
solutions $\gamma_i (\alpha)$ ($i=0,x,y$) if and only if the secular
condition is fulfilled:
\begin{equation}
\label{eq:secular}
\det \left[ \lambda(\alpha)\delta_{ik} -
\sum_{j=0,x,y} \eta_{ij} \langle U_j U_k \rangle \right]=0,
\end{equation}
which allows one to determine $\lambda(\alpha)$ as a function of the parameters
$u_0$, $u_x$, $u_y$ and of
the chemical potential $\mu$. These should be regarded as the eigenvalues
of the operator $\hat{V}_t$ at the beginning of the RG flow ($t=0$). Their
sign therefore accounts for an instability: the existence of a negative
eigenvalue $\lambda(\alpha)$ heralds a later divergence pattern for
$t\rightarrow t_c$, $t_c$ being determined by the most negative of the
$\lambda(\alpha)$. The eigenfunctions may then be determined by standard
algebra through the coefficients $\gamma_i (\alpha)$ of
their expansion in terms of the $U_i$.

We remark that the main numerical task is to evaluate,
once for all, the integrals defined in
Eq.~(\ref{eq:eigsysint}). The secular condition,
Eq.~(\ref{eq:secular}), can then be easily discussed
by changing at will the values of the coupling parameters.
Besides, the overall formalism allows a straightforward
generalization, by adding other terms in the development
of the potential function, Eq.~(\ref{eq:potentialbis}),
thus increasing the order of the matrix $\langle U_i U_j \rangle$.

\section{Discussion and final remarks}

The results we have exposed thus far may be discussed with respect to
the various possible values in the couplings, $u_0$, $u_x$, $u_y$.

In the case we may neglect an on-site pairing ($u_0 =0$),
Eq.~(\ref{eq:secular}) factorizes to yield:
\begin{eqnarray}
\lambda(0) && = 0\nonumber\\
\lambda(1,2) &&= (u_x +u_y )\langle U_x^2 \rangle
\pm \sqrt{(u_x -u_y )^2 \langle U_x^2 \rangle^2 +
4u_x u_y \langle U_x U_y \rangle^2} .
\end{eqnarray}

Furthermore, in the symmetric case, $u_x =u_y$, the non-vanishing
eigenvalues reduce to:
\begin{equation}
\label{eq:lambdasymm}
\lambda(1,2) = 2 u_x [\langle U_x^2 \rangle \pm \langle U_x U_y \rangle ],
\end{equation}
which are displayed in Fig.~\ref{fig:eigenvalues-intersite} as functions
of the chemical potential $\mu$.

Since $\langle U_x^2 \rangle \geq \langle U_x U_y \rangle$, for any
value of $\mu$, the eigenvalues Eq.~(\ref{eq:lambdasymm}) are both negative
for any attractive inter-site coupling $u_x <0$, however weak.
A superconducting ground state is thus predicted for any filling $N$ of the
band, although the critical temperature is undistinguishable from zero
even as $N\gtrsim 0.5$. This may be regarded as a mere consequence of
the very pronounced peak in the density of states (Fig.~\ref{fig:statedensa}),
which determines the behaviour of all the eigenvalues.

As shown in Fig.~\ref{fig:Tc-intersite}, insertion of the most negative
eigenvalue into Eq.~(\ref{eq:Tc}) yields a maximum for $T_c$, and the
existence of a superconducting phase for $N\lesssim 0.5$, with $u_x =-1.0$.
Here, the occurrence of one or more peaks is not a serious problem in
comparison with the experimental results, which predict a smooth plateau
for $T_c$ \emph{vs} $N$:~\cite{Zhang} in fact, the experimental data
are relative to different samples with different composition, and any
eventual peak would be smeared out on the average. The trend in $T_c$
\emph{vs} $N$ predicted by this RG analysis basically confirms the
analogous result recovered via direct solution of the standard mean-field
gap equation,~\cite{Spathis} even if $T_c$ is never rigorously zero,
except at the band edges. At variance with the experimental evidence,
for $u_0 =0$, $T_c$ has a steep increase at the band bottom, starting exactly
at $N=0$. We shall comment on this fact later on.

In the case of such a symmetric interaction, $u_x =u_y$, the exchange
symmetry $x\leftrightarrow y$, which is present in the hole dispersion
relation $\ve(\bk)$ Eq.~(\ref{eq:dispersionrelation}) and therefore
in the Fermi surface, is restored even for the interaction. Thus, the kernel
$V(\bth,\bth^\prime )$ is now symmetric under the exchange
$\theta_x \leftrightarrow \theta_y$,
$\theta_x^\prime \leftrightarrow \theta_y^\prime$. The eigenfunctions of
the integral operator $\hat{V}$ are even or odd with respect to such an
invariance transformation, and, as discussed at the end of Section~III,
the scale invariant gap equation (\ref{eq:gapformal}) admits even
($s$ wave) or odd ($d$ wave) solutions for the gap function.

Moreover, if $u_0 =0$, there are just two eigenfunctions in the
expansion for the gap, Eq.~(\ref{eq:gapexpanded2}), and at $T=0$
the gap equation admits the decoupled solutions:
\begin{equation}
\Delta (\bth) = {\mathrm const}~\Phi_1 (\bth);~~~
\Delta (\bth) = {\mathrm const}~\Phi_2 (\bth),~~~(T=0),
\end{equation}
where the proportionality constants are fixed by the scale $\Delta_0$
defined by Eq.~(\ref{eq:delta01}) and explicitly given by
Eq.~(\ref{eq:delta02}):
\begin{equation}
\label{eq:delta03}
\Delta_0 = 2 \Lambda_t \exp \left(\lambda_t^{-1} (1,2) \right).
\end{equation}

The eigenfunctions $\Phi_1$, $\Phi_2$ follow by direct solution
of the system Eq.~(\ref{eq:eigsys}), which yields $\gamma_y (1)=
\gamma_x (1)$, $\gamma_y (2) =-\gamma_x (2)$, whence:
\begin{eqnarray}
\Phi_1 (\bth) &&= {\mathrm const}~ [\cos\theta_x + \cos\theta_y ]
{}~~~~~\mbox{$s$ wave} ,\nonumber\\
\Phi_2 (\bth) &&= {\mathrm const}~ [\cos\theta_x - \cos\theta_y ]
{}~~~~~\mbox{$d$ wave} .
\end{eqnarray}
The latter equations, together with Eq.~(\ref{eq:delta03}), fix the
two symmetric solutions for the gap function. While in general a
broken-symmetry solution may occur (\emph{e.g.,} a mixed $s$-$d$
wave~\cite{Spathis}), a pure $s$ or $d$ wave solution is always expected
close to the transition points, which for $T=0$ are $N=0$ and $N\approx 0.5$.
In Fig.~\ref{fig:eigenvalues-intersite} we observe a cross-over around
the peaked region of the eigenvalues: at the bottom of the band, $\mu=\mu_1$,
the most negative eigenvalue corresponds to the even eigenfunction $\Phi_1$,
while for larger values of $N$ the most negative eigenvalue corresponds
to the odd eigenfunction $\Phi_2$. The cross-over explains the occurrence
of a double peak for $T_c$ in Fig.~\ref{fig:Tc-intersite}, and allows
us to predict, at the transition point, the opening of an $s$-wave gap
for $N\lesssim 0.2$, and of a $d$-wave gap for $N\gtrsim 0.2$. At $T=0$,
an intermediate broken-symmetry solution is awaited around the cross-over.
All this is in agreement with a previous mean-field analysis.~\cite{Spathis}

Now let us switch on an on-site interaction, $u_0 \neq 0$. In general,
the cubic equation (\ref{eq:secular}) may be solved analytically,
but it is instructive to look a little closer at the case of a symmetric
inter-site interaction, $u_x =u_y$. Again, the eigenfunctions of $\hat{V}$
must be even or odd with respect to the exchange $x\leftrightarrow y$.

The three linearly independent functions $U_0$, $U_x$, $U_y$
may be written in terms of the set $\{ U_0 ,U_\pm \}$, being
$U_\pm = [U_x \pm U_y ]/\sqrt{2}$. Since both $U_0$
and $U_+$ are even, then the function $U_-$, which is odd,
generates a one-dimensional invariant subspace for $\hat{V}$, so that
$U_- \equiv\Phi_2$ must be an eigenfunction for $\hat{V}$, and its eigenvalue
$\lambda(2)$ cannot be affected by the presence of an on-site coupling.
Therefore,
\begin{eqnarray}
\Phi_2 &&= {\mathrm const}~[\cos\theta_x -\cos\theta_x ],\\
\lambda(2) &&= 2u_x [\langle U_x^2 \rangle -\langle U_x U_y \rangle ].
\end{eqnarray}
In fact, in the symmetric case, the cubic equation (\ref{eq:secular})
factorizes as:
\begin{equation}
\lambda(0,1)= \frac{u_0}{2} \langle U_0^2 \rangle
+ u_x [\langle U_x^2 \rangle + \langle U_x U_y \rangle ]
\pm\sqrt{\left[ \frac{u_0}{2} \langle U_0^2 \rangle
-u_x \left( \langle U_x^2 \rangle + \langle U_x U_y \rangle \right)
\right]^2 + 4u_0 u_x \langle U_0 U_x \rangle^2}.
\end{equation}

About the choice of the parameters $u_0$, $u_x$, we may
anticipate that the overall trend for $T_c$ \emph{vs} $N$ is not relevantly
affected by any reasonable change in their values, since it is a direct
consequence of the chosen dispersion relation, as it is quite evident
by comparison with the density of states pictured in
Fig.~\ref{fig:statedensa}. Nonetheless, the most interesting scenario
shows up when there is a competition between on-site and inter-site
interactions. Namely, for $u_0 =-1.0$, $u_x =u_y =1.0$, the eigenvalues
are reported in Fig.~\ref{fig:eigenvalues-onsite}, and are characterized
by some nice features: first of all, one only eigenvalue, $\lambda(0)$,
is negative and leads the phase-transition; we may then notice that at
the band bottom, $N\rightarrow 0$, the negative eigenvalue drops to
a very small value at a finite filling $N>0$; finally, the same eigenvalue
goes to zero almost linearly inside the band for $\mu\approx 2.26$, where
a phase transition is expected even at $T=0$. Actually, the eigenvalue never
crosses the $\lambda=0$ axis, as it is shown in the insert of
Fig.~\ref{fig:eigenvalues-onsite}. The eigenvalues $\lambda(0)$ and
$\lambda(1)$, whose corresponding eigenfunctions share the same symmetry,
cannot cross and repel each other, as predicted by a general theorem due
to E. Wigner and J. von Neumann.~\cite{Landau} However, we notice that
a small negative eigenvalue is equivalent to a vanishing value since,
due to the exponential in Eq.~(\ref{eq:Tc}), the critical temperature
becomes extremely low and thus negligible.

Comparing with the case of a pure negative on-site interaction
($u_x =u_y =0$), when the eigenvalue coincides (up to a factor)
with the density of states $n$ of Fig.~\ref{fig:statedensa}, we
can assert that the presence of a non-zero nearest-neighbour repulsion
reduces the range of the superconductive phase inside the band. This
is not trivial since, for instance, if $u_x <0$, the presence of
an on-site repulsive interaction $u_0 >0$ does not produce any relevant
effect on the phase diagram. In fact, reversing the sign of all the
eigenvalues in Fig.~\ref{fig:eigenvalues-onsite}, and comparing with
Fig.~\ref{fig:eigenvalues-intersite}, we notice that the presence
of the $d$ wave eigenvalue $\lambda(2)$, whose value is not affected
by $u_0$, largely reduces the weight of an on-site repulsion against
the inter-site attractive coupling.

The progressive reducing of the superconducting phase with the increase
of the inter-site repulsion is illustrated in
Fig.~\ref{fig:eigenvalues-polyonsite} and Fig.~\ref{fig:Tc-polyonsite}.
Here, $u_0 =-4.5$ and $u_x$ varies from $0.5$ to $3.5$.
Fig.~\ref{fig:eigenvalues-polyonsite} displays the negative eigenvalue
while the critical temperature is reported in Fig.~\ref{fig:Tc-polyonsite}.
We notice that, as a consequence of the competition between $u_0$ and
$u_x$, now at the bottom of the band the critical temperature drops to
zero at a finite filling $N\gtrsim 0$, ranging from $0$ to $0.1$, in
qualitative agreement with the experimental data.~\cite{Zhang}

At the critical point the gap function is described by the even
eigenfunction:
\begin{equation}
\Phi (\bth) = \gamma_0 + \gamma_x (\cos\theta_x +\cos\theta_y ),
\end{equation}
being $(\gamma_0 ,\gamma_x ,\gamma_y =\gamma_x )$ the solution of the linear
equations (\ref{eq:eigsys}).

Up to now we did not make any effort in order to justify
the physical origin of the interaction Eq.~(\ref{eq:potentialbis}).
This simple RG analysis seems to suggest the occurrence of a competition
between an on-site negative Hubbard coupling and a repulsive nearest-neighbour
interaction, on the basis of the comparison with the experimental data.
We notice that, while several microscopic models predict the occurrence
of a short-range attractive coupling,~\cite{Micnas}
the nearest-neighbour repulsion
could be justified by the presence of a long-range Coulomb interaction.
Nonetheless, we must caution, since for a full analysis of the pairing
interaction a larger number of terms should be retained in its
definition
Eq.~(\ref{eq:potential}), thus increasing the order of the coupling
matrix, Eq.~(\ref{eq:matrix}).

\acknowledgements

We would like to thank prof. R. Shankar for encouraging suggestions
and Dr. D. Zappal\`a for friendly discussions.
One of us (GGNA) enjoys a CNR/GNSM scholarship.

\begin{figure}
\caption{The `tree' and the `tadpole' diagrams are shown, which contribute
to the interaction part of the action,
Eq.~(\protect\ref{eq:interactionaction}),
at a tree level.}
\label{fig:treelevel}
\end{figure}

\begin{figure}
\caption{We show the three second-order one-loop diagrams, together
with the factors arising from symmetry and statistics requirements.
The only one which gives a relevant contribution is the {\sc bcs}
one.~\protect\cite{Shankar}}
\label{fig:onelooplevel}
\end{figure}

\begin{figure}
\caption{RG flow diagram. The flow lines for $(\beta\Lambda,\lambda)$ have been
numerically evaluated for different values of the unrenormalized eigenvalue
$\lambda_0$. The inverse temperature $\beta$ is accordingly seen to
renormalize either to zero or to some finite value, as soon as $\lambda_t$ is
able to diverge at a finite $t=\bar{t}$.
The dashed line joins the critical points for the different values
of $\lambda$, and tends asymptotically to the axis $\beta=0$ as
$t\rightarrow\infty$.}
\label{fig:flow}
\end{figure}

\begin{figure}
\caption{The Fermi surfaces corresponding to a hole dispersion relation
Eq.~(\protect\ref{eq:dispersionrelation})
are shown, in correspondence to increasing values of the chemical potential,
$\mu=-2.1,~-1.8,~0.0,~3.0,~5.8$. The box selects only the positive octant
of the first Brillouin zone ($0\le k_x ,k_y ,k_z \le \pi$),
being $\Gamma=(0,0,0)$, $\bar{M} = (\pi,0,0)$, $X=(\pi,\pi,0)$,
$Y=(\pi,\pi,\pi)$.}
\label{fig:FS}
\end{figure}

\begin{figure}
\caption{The density of states $n(\mu)$,
Eq.~(\protect\ref{eq:densityofstates}),
is shown for the normal (non-interacting) Fermi system, as
function of the chemical potential $\mu$,  $\mu_1 \le\mu\le \mu_3$.
It displays a pronounced, yet finite, maximum for
$\mu=\mu_2 =-2.0$. It is a token of the quasi-bidimensionality of the
system, due to its layered structure, and would have been a real van Hove
singularity if no inter-plane hopping had been considered ($C=0$).}
\label{fig:statedensa}
\end{figure}

\begin{figure}
\caption{The total fraction of occupied states $N(\mu)$,
Eq.~(\protect\ref{eq:occupiedstates}),
is shown for the normal (non-interacting) Fermi system, as
functions of the chemical potential $\mu$,  $\mu_1 \le\mu\le \mu_3$.
The normalization condition at the top of the band
$N(\mu_3 )=1$ is clearly fulfilled.}
\label{fig:statedensb}
\end{figure}

\begin{figure}
\caption{The two non-zero eigenvalues $\lambda(1)$ (dashed line) and
$\lambda(2)$ (full line) are
here displayed over $\mu$, $\mu_1 \leq\mu\leq\mu_3$,
for the values of the parameters $u_0 =0.0$,
$u_x =u_y =-1.0$.}
\label{fig:eigenvalues-intersite}
\end{figure}

\begin{figure}
\caption{We show here the ratio $T_c /\Lambda$
for the values of the parameters $u_0 =0.0$,
$u_x =u_y =-1.0$, as a function of the fraction of
occupied states $N$.}
\label{fig:Tc-intersite}
\end{figure}

\begin{figure}
\caption{We show the eigenvalues $\lambda(0)$ (negative),
$\lambda(1)$ (positive, dashed) and $\lambda(2)$ (full line)
over $\mu$, $\mu_1 \leq\mu\leq\mu_3$,
in the symmetrical case corresponding to
the values of the parameters $u_0 =-1.0$,
$u_x =u_y =1.0$. The detail focuses on the region around $\mu\simeq 2.262$,
where the two eigenvalues corresponding to the same symmetry ($\lambda(0)$
and $\lambda(1)$) run closely without crossing. The eigenvalue $\lambda(0)<0$
is seen to generate an instability.}
\label{fig:eigenvalues-onsite}
\end{figure}

\begin{figure}
\caption{We show the only eigenvalue giving rise to an instability,
$\lambda(0)$, as a function of $\mu$, $\mu_1 \leq \mu\leq\mu_3$,
for the values of the parameters (from bottom to top): $u_0 =-4.5$,
$u_x =u_y =0.5,~1.5,~2.5,~3.5$.}
\label{fig:eigenvalues-polyonsite}
\end{figure}

\begin{figure}
\caption{We show the ratio $T_c /\Lambda$ as a function of the
fraction of occupied states $N(\mu)$,
for the values of the parameters (from top to bottom): $u_0 =-4.5$,
$u_x =u_y =0.5,~1.5,~2.5,~3.5$.}
\label{fig:Tc-polyonsite}
\end{figure}
\end{document}